\documentclass[aps,prx,twocolumn,groupedaddress,nofootinbib]{revtex4-2}

\usepackage{microtype}
\usepackage{amsmath}
\usepackage{amssymb}
\usepackage{amsfonts}
\usepackage{slashed}
\usepackage{mathtools}
\usepackage{xcolor}
\usepackage{ninecolors}
\usepackage{orcidlink}
\usepackage{hyperref}
\hypersetup{
  unicode,
  pdftitle={The instanton sum in two-dimensional self-dual models},
  pdfauthor={Mendel Nguyen},
  pdfprintscaling=None,
  breaklinks,
  colorlinks=true,
  linkcolor=azure5,
  citecolor=azure5,
  filecolor=magenta,
  urlcolor=blue5,
}

\newcommand{\del}{\partial}
\DeclareMathOperator{\Tr}{Tr}

\begin{document}

\title{The instanton sum in two-dimensional self-dual models}

\author{Mendel Nguyen}
\affiliation{Centre for Particle Theory, Department of Mathematical Sciences, Durham University, Durham DH1 3LE, UK}

\begin{abstract}
This paper considers the \emph{self-dual $CP^{N-1}$ model} in two dimensions, which is defined by restricting the path integral of the ordinary two-dimensional $CP^{N-1}$ model to holomorphic configurations.
We find that the instanton measure is one-loop exact and agrees with the one-loop instanton measure in the ordinary model. 
For the $N=2$ case, the instanton gas admits a description in terms of free massive Dirac fermions.  
Nevertheless, we find that the self-dual model is gapless.
We also comment on the $\theta$ dependence. 
\end{abstract}

\maketitle

\section{Introduction}

While instantons explain many important nonperturbative effects in field theory, they have so far not succeeded in explaining either the mass gap or confinement properties of pure Yang--Mills theory in four dimensions.
The essential difficulty is that summing over increasingly big instantons generates an infrared divergence, indicating the breakdown of the semiclassical expansion. 

But the instanton physics might be tractable in a world where non-self-dual configurations (meaning configurations with $F \neq *F$ in Euclidean space) have zero weight in the path integral.
Such a world is described by \emph{self-dual Yang--Mills theory} \cite{Chalmers:1996rq}, which has long been a productive setting for perturbative studies of QCD amplitudes \cite{Bardeen:1995gk, Cangemi:1996rx, Witten:2003nn}.
That it might also be a promising setting for nonperturbative studies \cite{Frenkel:2006fy, Frenkel:2008vz} is hinted by the fact that there the quantum instanton measure is one-loop exact (and agrees with the one-loop instanton measure in full Yang--Mills theory) \cite{Losev:2017qrj, Bittleston:2025jmk}. 

Further, it could be advantageous to study the full theory by approaching it from the self-dual one. 
Ref. \cite{Domurcukgul:2025xgf} suggested a way to do this by gradually reincorporating the effects of non-self-dual configurations through a one-parameter deformation of the action that continuously connects the self-dual theory to the full one.
Thus, a plausible program for tackling full Yang--Mills theory is to first sum up instanton effects in the self-dual theory, and then devise a controlled approximation to study the continuous deformation towards the full theory.

The purpose of this note is to test out the feasibility of this program by following the time-honored tradition of warming up in the simpler setting of two-dimensional nonlinear sigma models.
Specifically, we consider the two-dimensional $CP^{N-1}$ model, which, like four-dimensional Yang--Mills theory, has asymptotic freedom, instanton configurations, and a mass gap (which still has not been analytically established outside the $1/N$ expansion) \cite{DAdda:1978vbw, Witten:1978bc}. 
Also in complete analogy with the Yang--Mills case, one may construct a self-dual $CP^{N-1}$ model (to be defined below), as well as a one-parameter deformation connecting it to the full model \cite{Hayashi:2026tgo}.

Here, our attention is confined to just the first step of summing up instantons in the self-dual model.
We will see that just as in the case of Yang--Mills theory, the quantum instanton measure in the self-dual $CP^{N-1}$ model is one-loop exact and coincides with the one-loop measure in the full model. 
For the $N=2$ case ($CP^1$), we will be able to find the exact instanton sum for the partition function by simply repurposing a well-known one-loop computation of Fateev, Frolov and Schwarz \cite{Fateev:1979dc}.
We have not done the exact instanton sum for general $N$, but we will still be able to comment on the $\theta$ dependence.
Finally, we will see that despite the proliferation of instantons, the self-dual model is \emph{exactly gapless}, suggesting that the interactions between self-dual and anti-self-dual instantons are essential for generating the mass gap in the full model.

\section{Full \boldmath{$CP^1$} model}

We will start with a review of some basic features of the full $CP^1$ model in two-dimensional Euclidean space.
The model can be formulated in terms of a single complex field $w$ with action given by
\begin{equation}
    S = \frac{4}{g^2} \int d^2 x (\lvert \del_+ w \rvert^2 + \lvert \del_- w \rvert^2)/\rho^2
\end{equation}
where $\rho = 1+|w|^2$, $g$ is a coupling constant, and we are using complex coordinates $x^\pm = x^1 \pm i x^2$ (which below will also be denoted by $x=x^+,\bar x=x^-$).  
Finite-action configurations are sorted into sectors labeled by the instanton number
\begin{equation}
    Q = \frac{1}{\pi} \int d^2 x (\lvert \del_+ w \rvert^2 - \lvert \del_- w \rvert^2)/\rho^2
\end{equation}
which can only take integer values. 
The partition function is
\begin{equation}
    Z = \int D\mu[w] \; e^{-S + i \theta Q}
\end{equation}
where the path integral measure is formally given by
\begin{equation}
    D\mu[w] \equiv \prod_x \frac{d^2 w(x)}{g^2 \rho^2(x)}
\end{equation}
The partition function can be written as the Fourier sum
\begin{equation}
    Z = \sum_{Q = -\infty}^\infty Z_Q e^{i\theta Q}
\end{equation}
with $Z_Q$ denoting the path integral of $e^{-S}$ over the instanton-number $Q$ sector. 

It is easy to see that 
\begin{equation}
    S = \frac{\pm 4\pi Q}{g^2} + \frac{8}{g^2} \int d^2 x \lvert \del_\mp w \rvert^2/\rho^2
    \label{eq:CP1-action-fluctuation}
\end{equation}
Thus, in any sector of fixed instanton-number, the action is least on the holomorphic configurations if $Q\geq0$, or on the anti-holomorphic configurations if $Q\leq0$, and in either case the least action is $4\pi \lvert Q \rvert/g^2$.
Below, we shall also say `self-dual' for `holomorphic'. 
The self-dual configurations of instanton number $Q\geq0$ are given by the rational holomorphic functions of order $Q$,
\begin{equation}
    w(x) = c\prod_{r=1}^Q \frac{x-a_r}{x-b_r}
    \label{eq:sd-config}
\end{equation}
where $a_1,b_1,\ldots,a_Q,b_Q,c$ are any complex numbers.
We shall impose the boundary condition $w(x) \to 1$ as $|x| \to \infty$ so that we always have $c=1$.
The anti-self-dual configurations are complex conjugates of self-dual configurations. 
We denote the moduli space of instanton-number $Q$ self-dual or anti-self-dual configurations by $\mathcal M_Q$. 

Let us now consider fluctuations around a self-dual configuration. 
We can write a generic field configuration in the form
\begin{equation}
     w = w_{\rm sd} + g \rho_{\rm sd} \xi
\end{equation}
where $w_{\rm sd}$ is a self-dual configuration as in \eqref{eq:sd-config}, $\rho_{\rm sd} = 1 + \lvert w_{\rm sd} \rvert^2$, and $\xi$ is a new variable.%
\footnote{
The reason we extract a factor $\rho_{\rm sd}$ from the deviation $w - w_{\rm sd}$ is so that the natural functional norm of $\xi$ is given by $\int d^2 x \xi^* \xi$.
}
Substituting this into the action, and using \eqref{eq:CP1-action-fluctuation}, we see that 
\begin{equation}
    S = \frac{4\pi Q}{g^2} + 8 \int d^2 x \frac{ \lvert \del_-(\rho_{\rm sd} \xi) \rvert^2}{\rho_{\rm sd}{}^2} + O(\xi^3)
    \label{eq:quadratic-action}
\end{equation}
Requiring $\xi$ to be orthogonal to fluctuations in $w_{\rm sd}$, we may change path integration variables from $w$ to $w_{\rm sd},\xi$. 
This transforms the path integral measure
into $d\mu_0 D\xi$, with $d\mu_0$ denoting the ``classical'' measure for integrating over the moduli space $\mathcal M_Q$, i.e. for integrating over the $a$s and $b$s which parametrize the self-dual configurations \eqref{eq:sd-config}, and using the approximation where the distinction between $\rho_{\rm sd}$ and $\rho$ in the Jacobian is ignored.

Doing the path integral over $\xi$ in the Gaussian approximation then gives the one-loop quantum measure $d\mu_{\rm 1loop}$ on the instanton moduli space. 
Fateev, Frolov, and Schwarz (FFS) \cite{Fateev:1979dc} (see also Ref. \cite{Berg:1979uq}) were first to do this, and they found the elegant result
\begin{multline}
    d\mu_{\rm 1loop} \equiv d\mu_0 \int D \xi \; (e^{-S})_{\rm Gauss}
    \\ = {\rm const.} \frac{\Lambda^{2Q} e^{-4\pi Q/g^2}}{g^{4Q}(Q!)^2} \prod_{i=1}^Q d^2 a_i d^2 b_i \biggl\lvert \det \biggl( \frac{1}{a_i-b_j} \biggr) \biggr\rvert^2
    \label{eq:FFS}
\end{multline}
where $\Lambda$ is an ultraviolet cutoff, and the proportionality constant is independent of $Q$.
Further, they recognized that the result of integrating over the moduli space $\mathcal M_Q$ (i.e. over all $a$s and $b$s) and summing over all nonnegative integer $Q$ is literally the partition function of free massive Dirac fermions,
\begin{equation}
    Z_{\text{free Dirac}} = \int D\psi D\bar\psi \exp\biggl\lbrack \int d^2 x \; \bar \psi (\slashed \del + M) \psi \biggr\rbrack
    \label{eq:free-Dirac}
\end{equation}
with
\begin{equation}
    M = {\rm const.} \Lambda e^{-2\pi/g^2} / g^4
    \label{eq:FFS-mass}
\end{equation}

This computation is quite striking.
It shows that while the moduli space integrals are each infrared divergent, their sum is harmless; they get reinterpreted as the typical infrared divergences one gets from expanding a massive theory around a massless one. 
It also seems to show that instanton proliferation has generated a mass gap. 

But striking as the FFS computation is, it does not make sense as a computation of anything in the full $CP^1$ model, as it totally neglects the contributions from the negative instanton-number sectors.
Moreover, we cannot even take the integral of \eqref{eq:FFS} over $\mathcal M_Q$ as a reliable approximation to $Z_Q$ (for $Q\geq 0$).
For one, it is unclear to what extent higher-loop corrections, which are inevitably large, would change the answer. 
But more importantly, the computation entirely neglects ``instanton--anti-instanton'' configurations, i.e. configurations of the form $w \approx w_{\rm sd} w_{\rm asd}$ with $w_{\rm sd}$ self-dual and $w_{\rm asd}$ anti-self-dual. 
Neglect of such configurations is not justified.

But suppose we define a new model by throwing out all non-self-dual configurations.
Could it be that what FFS computed is actually meaningful in such a model?

\section{Self-dual \boldmath{$CP^1$} model}

We get the partition function of the self-dual model from that of the full model by inserting a delta functional under the path integral setting $\del_- w = 0$:
\begin{equation}
    Z = \int D\mu[w] \; \delta[\del_- w / g \rho] \;  e^{-S + i \theta Q} 
\end{equation}
Note that we have put $\del_-w/\rho$ inside the delta functional rather than simply $\del_-w$ (or any other expression proportional to $\del_-w$) in order for the self-dual model to realize $SO(3)$ symmetry nonlinearly as in the full model.%
\footnote{
Under an $SO(3)$ transformation, $\del_\mu w / \rho$ changes by a phase factor, so $\delta [\del_- w / \rho]$ is $SO(3)$ invariant.
}

We now proceed to transform this expression for the partition function as follows. 
First, we may replace $S$ by $4\pi Q/g^2$ in the action since $\del_- w = 0$.
Second, we may rewrite the delta functional as a Fourier path integral over a `Lagrange multiplier' field $v$.
Thus,
\begin{multline}
    Z = \int D\mu[w]Dv \; e^{-4\pi Q/g^2 + i \theta Q}
    \\ \times \exp\biggl[ - \frac{1}{g} \int d^2 x \biggl( \bar v \frac{\del_-w}{\rho} - \frac{\del_+ \bar w}{\rho} v \biggr) \biggr]
\end{multline}

To compute the contribution from the instanton-number $Q$ sector, we make the substitutions
\begin{equation}
    w = w_{\rm sd} + g \rho_{\rm sd} \xi, \quad \bar v = \bar \eta \frac{\rho}{\rho_{\rm sd}} 
    \label{eq:fluc-var}
\end{equation}
and change path integration variables to $\xi,\eta$ and $w_{\rm sd}$.
The path integral measure $D\mu[w] Dv$ becomes $d\mu_0 D\xi D\eta$ with $d\mu_0$ as above.
The quantum instanton measure is $d\mu = d\mu_0 K$ where $K$ is the path integral over the fluctuations $\xi,\eta$:
\begin{equation}
    K = \int D\xi D\eta \exp\biggl\lbrack - \int d^2 x \biggl( \bar \eta \frac{\del_-(\rho_{\rm sd} \xi)}{\rho_{\rm sd}} 
    + \frac{\del_+(\rho_{\rm sd} \bar \xi)}{\rho_{\rm sd}} \eta \biggr) \biggr\rbrack
\end{equation}
Integrating the last term by parts, we can bring the integral in the exponent to look like the action of a massless Dirac field%
\footnote{
More properly, a K\"ahler--Dirac field.
But the distinction is unimportant in flat space.
}
minimally coupled to a background $U(1)$ gauge field:
\begin{equation}
    S_A = \int d^2 x ( \bar \eta D_- \xi + \bar \xi D_+ \eta)
\end{equation}
where $D_\mu = \del_\mu - i A_\mu$ is the covariant derivative relative to the gauge field background
\begin{equation}
    A_\mp = \pm i \del_\mp \rho_{\rm sd}/ \rho_{\rm sd}
\end{equation}
Thus, $K$ is simply the inverse primed determinant of the Dirac operator:
\begin{equation}
    K = \int D\xi D\eta \; e^{-S_A} = (\det\nolimits' \slashed D)^{-1}
    \label{eq:gauged-Dirac}
\end{equation}
(the inverse because $\xi,\eta$ are bosonic!). 

It is important to note that we have not made any approximations.
The measure on $\mathcal M_Q$ is \emph{exactly} $d\mu = d\mu_0 (\det\nolimits' \slashed D)^{-1}$. 
In other words, the self-dual model is one-loop exact. 

Moreover, this measure is exactly equal (apart from overall normalization) to the one-loop instanton measure in the full $CP^1$ model as computed by FFS.
To see this, note from \eqref{eq:quadratic-action} that the part of the action quadratic in $\xi$ can be written as the space integral of $\lvert D_- \xi \rvert^2$.
Thus, the one-loop instanton measure in the full $CP^1$ model is given by $d\mu_0$ times the inverse primed determinant of the operator $-D_+ D_- \propto -D^2 - i\tilde F$. 
The proportionality of the one-loop measures then follows from the well-known observation that the operators $-D^2 - i\tilde F$ and $\slashed D$ have the primed determinants.%
\footnote{
Since $\slashed D$ anticommutes with $\gamma_5$, a complete set of independent eigenfunctions for $\slashed D$ with nonzero eigenvalues can be taken to consist of a complete set of eigenfunctions $\psi_n$ with eigenvalues $i\lambda_n$ on the positive imaginary axis, together with the eigenfunctions $\gamma_5 \psi_n$ which have eigenvalues $-i\lambda_n$ on the negative imaginary axis.
Thus, $\det\nolimits'\slashed D = \prod_n \lambda_n{}^2$. 
On the other hand, $-(\slashed D)^2 = - D^2 - \gamma_5 i \tilde F$ commutes with $\gamma_5$.
Acting on the $\gamma_5=+1$ subspace, it clearly coincides with $-D^2 - i \tilde F$.
Thus, $\det\nolimits'(-D^2-i\tilde F)$ is a product over a complete set of eigenfunctions of $-(\slashed D)^2$ with $\gamma_5=+1$.
Such a set may clearly be taken to consist of the functions $(1+\gamma_5)\psi_n$, which have eigenvalues $\lambda_n{}^2$.
Thus, $\det\nolimits'(-D^2-i\tilde F)$ is also equal to $\prod_n \lambda_n{}^2$. 
}

Thus, it appears that the self-dual $CP^1$ model is precisely the model in which the calculation of FFS is not only meaningful but exact.
The partition function of the self-dual model is indeed exactly equal to the partition function \eqref{eq:free-Dirac} of the theory of free Dirac fermions
\begin{equation}
    M = \text{const.} \Lambda e^{-2\pi/g^2 + i \theta/2}
\end{equation}
taking into account the $\theta$ angle as well as a topological counterterm mentioned in Ref. \cite{Hayashi:2026tgo} which cancels the $g^{-4}$ factor in \eqref{eq:FFS-mass}.
Are we now to conclude that the self-dual model has a mass gap generated by instanton proliferation, and that it is exactly dual to the theory of free massive Dirac fermions? 

No. As announced in the introduction, the self-dual model is actually gapless.
The massive Dirac fermions do not capture all correlations in the model. 
We will now show this by exhibiting an exact power law correlation function.

The correlation function we shall consider will be the two-point functions of conserved currents associated with the internal nonlinearly realized $SO(3)$ symmetry.
It is simplest to consider the $U(1)$ subgroup of symmetries that is linearly realized, given by the subgroup of transformations $w \to e^{i\theta} w, v \to e^{i\theta} v$.
The associated conserved currents are given by
\begin{equation}
    J_+ = i \bar v w / g \rho, \quad J_- = i v \bar w /g\rho
\end{equation}
Mere current conservation reads $\del_- J_+ + \del_+ J_- = 0$, but it is easy to see from the equations of motion that we actually have the much stronger conditions
\begin{equation}
    \partial_- J_+ = 0, \quad \partial_+ J_- = 0
\end{equation}
which say that $J_+$ and $J_-$ are holomorphic and antiholomorphic, respectively.
This, together with Euclidean invariance, allows us to immediately infer the power law
\begin{equation}
    \langle J_\pm(x) J_\pm(y) \rangle \propto \frac{1}{(x^\pm - y^\pm)^2}
\end{equation}

We can actually go further and directly calculate these two-point functions to all orders in the instanton sum.
Let us first consider $\langle J_+ J_+ \rangle$.
In any given instanton-number sector, we can use the substitution \eqref{eq:fluc-var} to write
\begin{equation}
    J_+ = \bar \eta w_{\rm sd}/g\rho_{\rm sd} + \bar \eta \xi
\end{equation}
Let $\langle \mathcal O \rangle^A$ denote the average of an operator $\mathcal O$ over the fluctuations $\xi,\eta$ around a given self-dual background weighted by $e^{-S_A}/K$, so that the exact average of $\mathcal O$ is given by
\begin{equation}
    \langle \mathcal O \rangle = \frac{1}{Z} \sum_{Q=0}^\infty \int_{\mathcal M_Q} d\mu \langle \mathcal O \rangle^A 
\end{equation}
Wick contracting and noting that the only nonzero two-point functions are $\langle \xi \bar \eta \rangle^A$ and $\langle \eta \bar \xi \rangle^A$, we find
\begin{equation}
    \langle J_+ (x) J_+(y) \rangle^A = -\langle \xi(x) \bar \eta (y)\rangle^A \langle \xi(y) \bar \eta(x) \rangle^A
\end{equation}
Now $\langle \xi \bar \eta \rangle^A$ is a Green's function,
\begin{equation}
    D_- \langle \xi(x) \bar \eta (y) \rangle^A = \delta^2 (x-y)
\end{equation}
Using the fact that this Green's function is $1/\pi(x^+-y^+)$ when $D_- = \del_-$, and the fact that in the general case the covariant derivative is given by $D_- = \rho_{\rm sd}{}^{-1} \del_- ( \rho_{\rm sd} \cdot )$, it is easy to find that
\begin{equation}
    \langle \xi(x) \bar \eta(y) \rangle^A = \frac{\rho_{\rm sd}{}^{-1}(x) \rho_{\rm sd}(y)}{\pi(x^+-y^+)}
\end{equation}
But now we see that the factors of $\rho_{\rm sd}$ all cancel in $\langle J_+ J_+ \rangle^A$ which is therefore independent of the background,
\begin{equation}
    \langle J_+(x) J_+(y) \rangle^A 
    = \frac{1}{\pi^2(x^+-y^+)^2}
\end{equation}
Computing $\langle J_- J_- \rangle^A$ works in exactly the same way.
Thus, we find after summing over instantons simply 
\begin{equation}
    \langle J_\pm(x) J_\pm(y) \rangle = \frac{1}{\pi^2(x^\pm-y^\pm)^2}
\end{equation}

While the free massive Dirac fermion theory does not provide a full dual description of the self-dual $CP^1$ model, it still completely captures the $\theta$ dependence.
A standard calculation of the free Dirac partition function \eqref{eq:free-Dirac} gives
\begin{equation}
    \frac{1}{V} \ln Z = \frac{M^2}{4\pi} \biggl( \ln \frac{M^2}{\Lambda^2} + C \biggr)
\end{equation}
where $V$ is the spacetime volume and $C$ is a cutoff-dependent constant.
The presence of the cutoff shows that the $\theta$ dependence is not entirely universal.
This is a reflection of the logarithmic divergence in the contribution of small instantons to the partition function. 
Nevertheless, the $\theta$ dependence does certainly contain universal content. 
For instance, while the average topological charge density $\langle q \rangle = (\del/i\del \theta)(\ln Z/V)\rvert_{\theta=0}$ and the topological susceptibility $\chi = (\del/i\del\theta)^2 (\ln Z/V)\rvert_{\theta=0}$ are individually afflicted with cutoff dependence, their difference is not.
Indeed, we find the perfectly universal result
\begin{equation}
    \chi - \langle q \rangle = \lvert M \rvert^2/4\pi
\end{equation}

\section{Self-dual \boldmath{$CP^{N-1}$} models}

Here let us briefly outline how some of our analysis is generalized to the self-dual $CP^{N-1}$ model for any $N\geq 3$.

The $CP^{N-1}$ model can be thought of as a theory of $N-1$ complex scalar fields assembled into a vector $w$ with action given by
\begin{equation}
    S = \frac{4}{g^2} \int d^2 x (\del_- \bar w G \del_+ w + \del_+ \bar w G \del_- w)
\end{equation}
where $G = (G_{\bar r s})$ is an $(N-1)\times(N-1)$ matrix (the Fubini--Study metric) given by
\begin{equation}
    G_{\bar rs} = \frac{\del^2}{ \del \bar w^{\bar r} \del w^s} \ln( 1 + \lvert w \rvert^2)
\end{equation}
The instanton number
\begin{equation}
    Q = \frac{1}{\pi} \int d^2 x (\del_- \bar w G  \del_+ w - \del_+ \bar w G \del_- w)
\end{equation}
is integer-valued for finite-action configurations. 
In a sector of fixed $Q\geq 0$, the action is least on the holomorphic (self-dual) configurations, and the least action is $4\pi Q/g^2$. 

It is possible to choose a unitary frame $E$, i.e. an $(N-1)\times(N-1)$ matrix such that
\begin{equation}
    \bar E G E = 1
\end{equation}
that is a smooth function of $w$. 
The self-dual $CP^{N-1}$ model is then formulated by inserting $\delta [E^{-1} \del_- w/g]$ under the path integral. 
Equivalently, one can formulate the self-dual model by taking its partition function as
\begin{multline}
    Z = \int D\mu[w] Dv \; e^{-4\pi Q/g^2 + i \theta Q} 
    \\ \times \exp \biggl\lbrack - \int d^2 x (\bar v E^{-1} \del_- w - \del_+ \bar w \bar E^{-1} v) \biggr\rbrack
\end{multline}
with $v$ an $(N-1)$-component vector of complex scalar fields. 

In any given instanton-number sector, the substitution
\begin{equation}
    w = w_{\rm sd} + g E_{\rm sd} \xi, \quad \bar v = \bar \eta E_{\rm sd}{}^{-1} E
\end{equation}
exactly transforms the path integral measure $D\mu[w] Dv$ into $d\mu_0 D\xi D\eta$. 
We then find that the quantum instanton measure is given by $d\mu = d\mu_0 K$ where $K$ is as in \eqref{eq:gauged-Dirac} with $\xi,\eta$ now understood as complex $(N-1)$-vectors, and $A_\mu$ now understood as the $U(N-1)$ gauge field background
\begin{equation}
    A_- = \pm i E_{\rm sd}{}^{-1} \del_\pm E_{\rm sd}
\end{equation}
An entirely analogous argument shows that the quantum instanton measure $d\mu$ here is exactly the same as the one-loop measure in the full $CP^{N-1}$ as first computed in Ref. \cite{Berg:1979uq}. 

Gaplessness of the self-dual $CP^{N-1}$ model is established by looking at the conserved currents associated with the $U(N-1)$ subgroup of the nonlinearly realized $U(N)/U(1)$ symmetry that is linearly realized.
These currents are given by
\begin{equation}
    J_+^a = i v E^{-1} T^a w, \quad J_-^a = i \bar w T^a \bar E^{-1} v
\end{equation}
where the $T^a$ are a set of Hermitian generators of this $U(N-1)$ subgroup. 
As before, the $J_+^a$ are holomorphic and the $J_-^a$ are anti-holomorphic, and an entirely similar computation gives the exact result
\begin{equation}
    \langle J_\pm^a(x) J_\pm^b(x) \rangle = \frac{\Tr (T^a T^b)}{\pi^2 (x^\pm - y^\pm)^2}
\end{equation}
showing that the self-dual $CP^{N-1}$ model is indeed gapless.

We have not explicitly computed the instanton sum for the partition function in a closed form.
Nevertheless, its form can be deduced on general grounds.
From the one-loop exactness of the $\beta$ function, which coincides with that of the full $CP^{N-1}$ model, shows that the instanton sum is a series in
\begin{equation}
    \Lambda^N e^{-4\pi/ g^2 + i \theta}
\end{equation}
with $\Lambda e^{-4\pi/g^2N}$ a renormalization-group invariant mass scale. 
On dimensional grounds, assuming extensivity of the free-energy, we must have $\ln Z/V \propto \Lambda^2$. 
But this then suggests the $\theta$ dependence $\ln Z/V \propto e^{i2\theta/N}$. 
This is in fact compatible with the large $N$ analysis of Ref. \cite{Hayashi:2026tgo}.

\section{Conclusion}

We have said that in complete analogy with the Yang--Mills case \cite{Domurcukgul:2025xgf}, the self-dual $CP^{N-1}$ model may be given a one-parameter deformation that connects it to the full $CP^{N-1}$ model \cite{Hayashi:2026tgo}. 
Having seen that the self-dual $CP^{N-1}$ model is gapless, we should next ask whether the gaplessness persists for some finite range of the deformation or whether a gap forms after even an infinitesimal deformation.
In the former case, there should then be a phase transition, as the full model is known to be gapped; 
it would then be interesting to investigate the transition point.
In the latter case, the question to ask is whether the gap stays open as one cranks up the deformation parameter all the way until the full theory is reached.
If these questions can be answered, we would perhaps be in a better position to answer the same questions in the Yang--Mills case.

\begin{acknowledgments}
    The author thanks Tin Sulejmanpasic and Mithat \"Unsal for useful discussions, and Gregory Johnson for comments on the draft. 
    This work was supported by STFC number UKRI4226. 
\end{acknowledgments}

\bibliography{sdnlsm}

\end{document}